\documentclass[10pt, two column]{article}

\usepackage[affil-it]{authblk} 
\usepackage{etoolbox}
\usepackage{lmodern}
\usepackage{graphicx}
\usepackage{lipsum}
\usepackage[table,xcdraw]{xcolor}
\makeatletter
\patchcmd{\@maketitle}{\LARGE \@title}{\fontsize{18}{19.2}\selectfont\@title}{}{}
\makeatother

\newcommand\blankfootnote[1]{%
  \let\thefootnote\relax\footnotetext{#1}%
  \let\thefootnote\svthefootnote%
}

\usepackage[T1]{fontenc}
\usepackage[utf8]{inputenc}
\usepackage{mathptmx}
\usepackage{blindtext}
\usepackage{nopageno}
\providecommand{\keywords}[1]{\textbf{\textit{Keywords---}} #1}
\usepackage{geometry}
\title{\bf Discovery of complex anomalous patterns of sexual violence in El Salvador}
\author{{\bf Maria De-Arteaga$^{1,2,3}$*, Artur Dubrawski$^{1,2,3,4}$} \\
Auton Lab$^1$, Machine Learning Department$^2$, Heinz College$^3$, Robotics Institute$^4$\\
Carnegie Mellon University \\ 
mdeartea@andrew.cmu.edu
\vspace{2.5em}
}

\date{\vspace{-5ex}}

\begin{document}

\maketitle
\vspace{6em}

\begin{abstract}
When sexual violence is a product of organized crime or social imaginary, the links between sexual violence episodes can be understood as a latent structure. With this assumption in place, we can use data science to uncover complex patterns. In this paper we focus on the use of data mining techniques to unveil complex anomalous spatiotemporal patterns of sexual violence. We illustrate their use by analyzing all reported rapes in El Salvador over a period of nine years. Through our analysis, we are able to provide evidence of phenomena that, to the best of our knowledge, have not been previously reported in literature. We devote special attention to a pattern we discover in the East, where underage victims report their boyfriends as perpetrators at anomalously high rates. Finally, we explain how such analyzes could be conducted in real-time, enabling early detection of emerging patterns to allow law enforcement agencies and policy makers to react accordingly.\blankfootnote{Discussion paper, Data for Policy 2016 - Frontiers of Data Science for Government: Ideas, Practices and Projections. Cambridge, United Kingdom, September 2016. }



\end{abstract}

\keywords data mining; anomaly detection; pattern discovery; data visualization; sexual violence; El Salvador.

 
\section{Introduction}

The design of efficient policies requires a profound understanding of the phenomena it deals with. Data constitutes an invaluable source to gain such understanding, but if misused data can become obsolete or even misleading. In the case of sexual violence, data is often used at a micro level to conduct investigations by law enforcement agents, and it is also used at a macro level to produce general descriptive statistics. In this paper, we attempt to bridge the gap between these levels of analysis, using data science to uncover latent structures that emerge when sexual violence episodes are not independent from each other. Dependencies occur in the presence of driving forces such as organized crime or social imaginary. These phenomena establish links between criminal episodes that can be uncovered through data mining. In this paper we focus on two levels of analysis. First, a bivariate analysis through pivot table heat maps allows us to answer questions that correspond to conditional distributions, such as who are the main perpetrators conditioned on age or location. Second, we focus on emerging spatiotemporal anomalous patterns, which can guide policy makers to points in time when frequencies of specific types of crime are rising and react accordingly. For example, a detective in a municipality might receive five rape reports that took place in the victim's house, and even though it is an increase from the average of two such cases per week, it can be easily attributed to a fluke. However, if the detective knew such increase also occurred in four neighboring municipalities, he/she would notice an emerging pattern. We propose a way of finding such systematically emerging anomalous patterns through the use of an efficient data structure that allows us to automatically perform massive multivariate queries and report results that present a significant deviation from the expected behaviour.  

Our approach consists of using relatively simple data --records of reported rapes for which only six attributes are available-- discovering complex anomalous patterns hidden in it, and using data visualization to present identified patterns in a way that is easy for practitioners to understand. The key assumption in our analysis is that at least a portion of sexual violence episodes are linked to organized crime, social imaginary, or other latent structure, as opposed to being completely isolated events with no common causes. 

El Salvador recently made it to the headlines around the globe as the murder capital of the world \cite{Mug2016} and the most violent peacetime country \cite{Pla2016}. \textit{Maras}--gangs-- and gang related violence are currently the primary challenge to peace in the region, threatening human rights and governments' stability \cite{JMR2009,Ara2005}. During this time, lethal violence against women and girls has positioned El Salvador at the first place of the infamous global ranking of female homicide rates \cite{Gen2015}. 
Gang rape initiation of females who join \textit{maras}, and the use of sexual violence by \textit{maras} as a weapon against enemies have been documented by both academics and journalists \cite{CaP2013,Hum2007}, while researchers have also pointed at the cultural legitimization of violence as a driving force of sexual violence in the country \cite{Hum2004}. 

Previous research and documentation of such phenomena in El Salvador allows us to posit an underlying latent structure among reported rapes. We aim to gain better understanding of such structure and identify emerging anomalous patterns that can be of interest to policy makers, presenting the results through data visualizations that are compelling and easy to understand by practitioners. We propose a way of implementing such anomaly discovery in real-time. Perhaps our most relevant finding, which to the best of our knowledge has not been previously discussed in literature, corresponds to evidence of a pattern in the East of the country, where victims between 12 and 14 years old (inclusive) report being raped by their boyfriends at significantly high rates, with specific points in time when this phenomenon has further escalated. In the remainder of this paper, Section \ref{rel_work} briefly reviews related work, Section \ref{data} introduces the data, Section \ref{meth} explains our methodology, Section \ref{res} follows with the results, and Section \ref{conc} presents the conclusion. 



\section{Related work}
\label{rel_work}

Sexual violence in El Salvador has been studied by multiple researchers \cite{Woo2009,Hum2004,Hum2007,SGW2008}. \cite{Woo2006} focused on the civil conflict that ended in 1992, points at El Salvador as a country where sexual violence was distinctly low compared to other cases, with the vast majority of incidents occurring in the early stages of the war and perpetrated by the state forces. This is perhaps one of the only times in literature where El Salvador is referenced for its relatively low prevalence of sexual violence. \cite{Hum2007} documents the systematic use of sexual violence by gangs both as part of their \textit{modus operandi} and as an initiation ritual, where women are subjected to gang rape, known as \textit{el trencito}, before joining a gang. Hume has also studied the cultural legitimization of violence   as an element of male gender identity in the general population \cite{Hum2004}, indicating it has led to the perception of sexual violence as a part of gender relations. The prevalence of child sexual abuse before age 15 has been studied in \cite{SGW2008}, where they conclude the most common perpetrators nationwide are neighbors or acquaintances and male family members. 

To the best of our knowledge, anomaly detection techniques have not been used in the past as a tool to study sexual violence. However, such techniques have been proposed for detection and prevention of crime waves and crime epidemics in general \cite{GoH2003}\cite{NeG2007}. Additionally, the use of machine learning to forecast recidivism of domestic violence incidents in particular households was proposed in \cite{BHS2005}. Such research, even though thematically related, differs from ours in that it deals with individual predictions rather than detection of systematic patterns.

The  T-Cube data structure, used in this paper to enable fast massive screening, was proposed in \cite{SMD2007} as a tool for fast retrieval and analysis of time series data. It has since been used to analyze large scale multidimensional spatiotemporal datasets, and it has proven to be useful in multiple surveillance and outbreak detection tasks, like monitoring food and agriculture safety \cite{DuZ2010} and detecting disease outbreaks \cite{WSD2010}. A user interface known as T-Cube Web Interface, which uses the T-Cube data structure and allows practitioners to visualize results and perform drill-down analysis in real-time, was presented in \cite{RMS2007}. 

\section{Data}
\label{data}

\begin{figure}[ht!]
\centering
\includegraphics[width=0.5\textwidth]{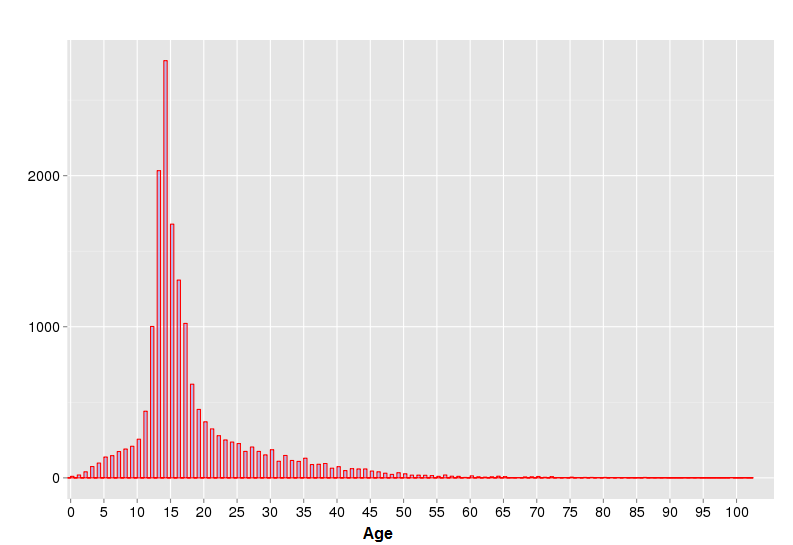}
\caption{Histogram of victim's age distribution.}
\label{age_plot}
\end{figure}

\begin{figure}[ht!]
\centering
\includegraphics[width=0.5\textwidth]{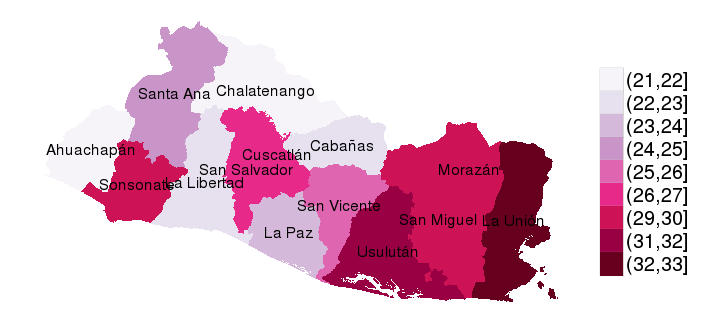}
\caption{Rate of total reported cases per 10,000 inhabitants.}
\label{map}
\end{figure}

The data used in this paper contains a record of all $16,965$ officially reported rapes between January 2006 and December 2014 in El Salvador\footnote{Official data provided by El Salvador's \textit{Instituto de Medicina Legal}.}\footnote{Data has been made publicly available and can be downloaded from http://especiales.laprensagrafica.com/2015/violaciones-en-elsalvador/}. For each case the exact date, age and gender of the victim, municipality and state where the rape took place, location (i.e. victim's house, empty lot) and relationship between the victim and the aggressor (i.e. father, acquaintance) are reported. In 15,739 cases the victim was female and in 1,225 the victim was male. The mean age of victims is 18.15, with a standard deviation of 9.76, and 7,595 victims are under fifteen years old. Figure \ref{age_plot} shows a histogram of age distribution, and Figure \ref{map} shows the rate of total reported rapes per 10,000 inhabitants for each state.  

\section{Methodology}
\label{meth}

Bivariate analyses through pivot table heat maps are used to visualize conditional distributions. Each row of the table represents the relative frequency of the column value conditioned on the row value, such that the sum across each row is 1. These tables give an overview where general trends and anomalies become visible.

Spatiotemporal anomaly detection is achieved through the use of the T-Cube data structure, which enables fast screening to detect those queries for which the observed counts deviate from the expected behaviour. An individual query is defined as the number of counts of a given event in a specified time window, where the following parameters are given:

\begin{enumerate}
\item Between one to three fixed attribute values.
\item Number of neighbouring locations to aggregate over, if one of the fixed attributes corresponds to a location.
\item An initial date for the time window.
\end{enumerate} 

Massive screening is defined as a search over all individual queries, where the parameters for the massive screening are the size of the time window and the list of attributes to query over. An example of a massive screening is a search over the attribute subset $(state,  age, perpetrator)$ for statistically significant time windows of seven days. Within this massive screening, an example of an individual query would be $(state=\{Morazan,SanMiguel,LaUnion\}, age=[12,14], perpetrator=boyfriend)$, for the week starting in $04-07-2014$.    

In massive screening, queries that significantly deviate from their expected counts are flagged as anomalies. Statistical significance tests are done using either Fisher's exact test \cite{Upt1992}, if the sample is small, or Chi-square test \cite{Eve1992}, if the sample is big. Both rely on the analysis of a contingency table, where we take into account the total count of events for the query's time window and for a reference window that illustrates past behaviour. 

\section{Results}

\begin{figure*}[ht!]
\centering
\includegraphics[width=0.88\textwidth]{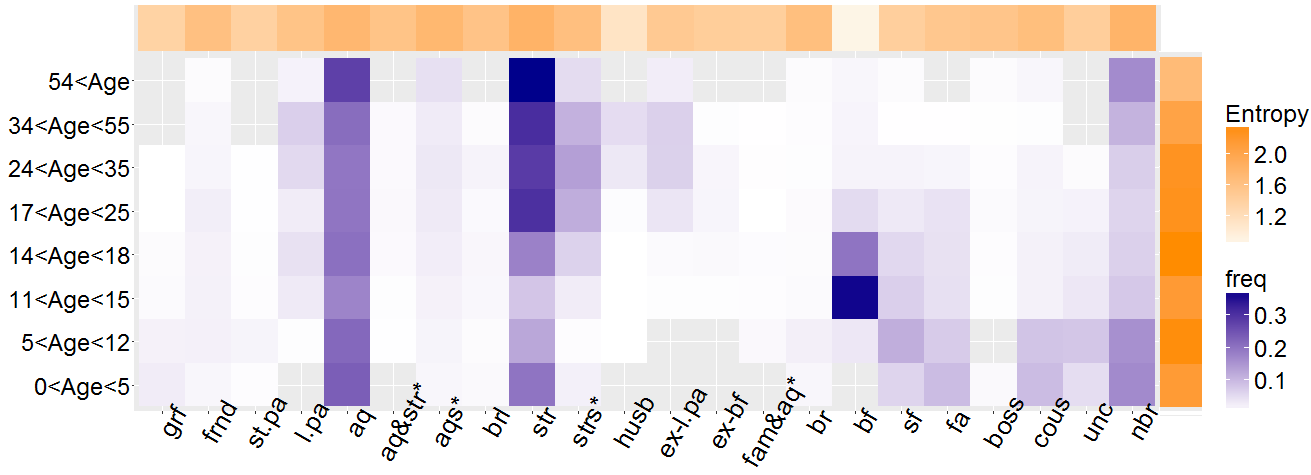}
\caption{Pivot table heat map of the distribution over perpetrator conditioned on victims' age range. Abbreviations: grf = grandfather; frnd = friend; st.pa = study partner; l.pa = life partner; aq = acquaintance; aq\&str = acquaintance \& strangers; aqs = acquaintances; brl = brother-in-law; str = stranger; strs = strangers; husb = husband;  ex-l.pa =ex-life partner; ex-bf = ex-boyfriend; fam\&aq = family members \& acquaintance; br = brother; bf = boyfriend; sf = stepfather; fa = father; boss = boss; cous = cousin ; unc = uncle; nbr = neighbour. ($^*$) indicates gang-rape.}
\label{heatmapage}
\end{figure*}

\begin{figure*}[ht!]
\centering
\includegraphics[width=0.88\textwidth]{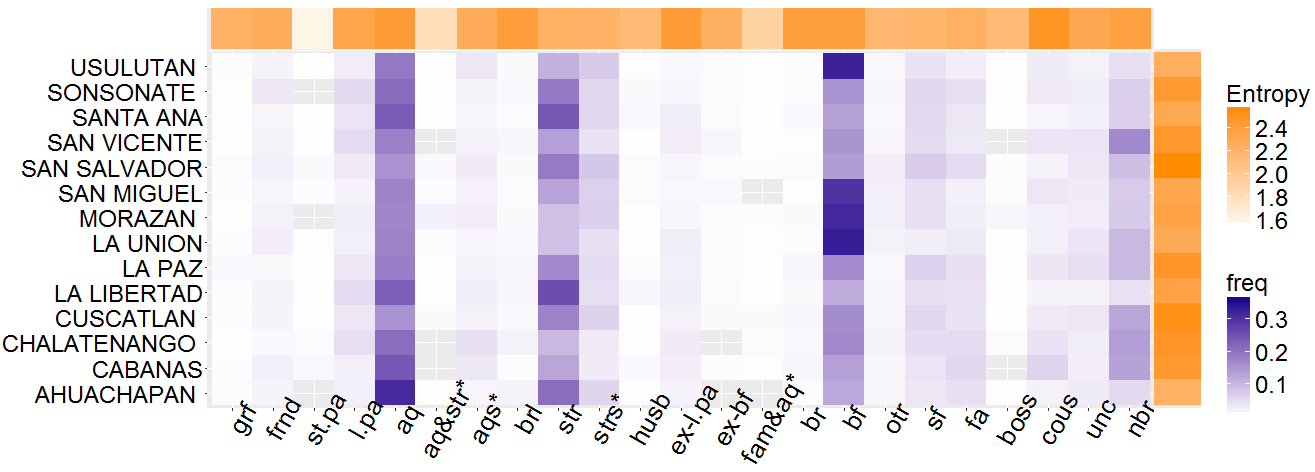}
\caption{Pivot table heat map of the distribution over perpetrators conditioned on state. For abbreviations see Figure \ref{heatmapage}.}
\label{heatmapstate}
\end{figure*}

Using pivot table heat maps, we visualize the conditional distribution over relationship between victim and aggressor conditioned on victims' age range, as well as conditioned over state. Figures \ref{heatmapage} and \ref{heatmapstate}, respectively, show the results. 

\label{res}

Looking at Figure \ref{heatmapage} we can analyse which aggressors are prevalent for each age group. Perhaps one of the most notable trends is that the most frequent aggressor of victims between 12 and 14 years old is the victim's boyfriend, which is not the case for other age ranges. Another interesting finding illustrated in this heat map is the fact that neighbours are responsible for a bigger proportion of rapes when victims are below 15 years old and above 55 years old. It is also relevant to note that, in line with findings in \cite{SGW2008}, strangers are responsible for a smaller portion of aggression against underage victims than adult victims.

Similarly, we study the most frequent aggressors per state, where we find that in Usulutan, San Miguel, Morazan and La Union the prevalence of reports where the boyfriend is the aggressor differs from that of the rest of the country. These are four neighbouring states in the East.

\begin{figure*}[ht!]
\centering
\includegraphics[width=0.7\textwidth]{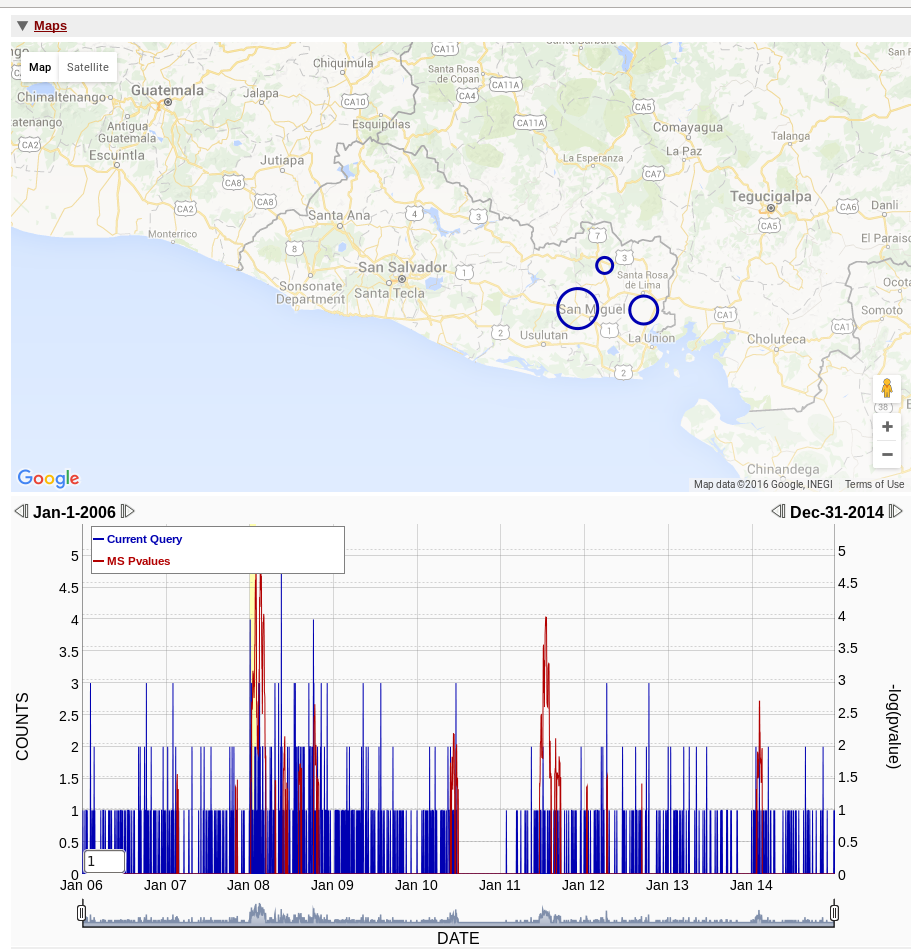}
\caption{Screenshot of TCWI. Timeline shows observed counts and the corresponding p-value of each query.}
\label{screenshot}
\end{figure*}

To gain insight into the patterns observed in the pivot tables, we perform a massive screening using T-Cube considering all queries where victims are between 12 and 14 years old and report their boyfriend as the aggressor. Each query compares the reported cases in a sub-region in a 28-days window, with the reported cases in the previous 365 days in that same area and the observed contemporary rates in the rest of the country. T-Cube allows us to screen in a matter of seconds over all queries of 28 consecutive days in all sub-regions of the country, considering aggregations of up to five neighbouring states with centroids no more than 50 kilometers away. The most highly ranked anomaly, with a p-value of 1.2$\times e^{-5}$, points at January 2008 in Morazan, La Union and San Miguel. Between January 3rd and January 30th that year, there were 23 reported rapes of this type, compared to an expected count of 6.93 per 28 days. Figure \ref{screenshot} shows a screenshot of this retrieved query. The period between May 3rd and May 30th of 2008 in Usulutan and San Miguel also appears as significantly anomalous, with an observed count of 21 such rapes compared to 6.57 expected. Table \ref{table_anomalies} shows the five highest ranking anomalies in this massive screening, for which 48 queries were flagged as anomalous with a p-value under 0.05.

\begin{table*}[ht!]
\centering
\label{my-label}
\begin{tabular}{|l|l|l|l|l|}
\hline
States                            & End date   & P-Value   & Count & Expected Count \\ \hline
\{LA UNION, MORAZAN, SAN MIGUEL\} & 01/30/2008 & 1.22E-05 & 23    & 6.93           \\
\{LA UNION, MORAZAN\}             & 02/18/2008 & 2.82E-05 & 17    & 4.38           \\
\{SAN MIGUEL, USULUTAN\}          & 05/30/2008 & 3.46E-05 & 21    & 6.57           \\
\{MORAZAN ,SAN MIGUEL\}           & 08/01/2011 & 3.74E-05 & 9     & 1.12           \\
\{SAN MIGUEL\}                    & 06/04/2008 & 7.04E-05 & 14    & 3.43           \\ \hline
\end{tabular}
\caption{Highest ranking anomalies for the massive screening with fixed attributes $age = [12,14]$,  $perpetrator = \{boyfriend\}$ and varying attribute $(States)$. Up to 5 states with centroids no more than 50 kilometers away are aggregated. The time window for each query is of 28 days, and the reference period is of 365 days.}
\label{table_anomalies}
\end{table*}

This method can also be used when we do not know a priori which subset of data will require attention. For example, when performing a massive screening over all possible values of location, state and aggressor, allowing for the anomalous cases to be determined by two or three of these attributes, we find that the highest ranked anomaly corresponds to the first 28 days of 2011, when there were 32 reported rapes that took place in a victim's house in the state of San Salvador, even though the expected count was 0.08. In the second place of this ranking we find that 63 rapes took place in a victim's house in the states of San Salvador and Cuscatlan between February 11th and March 11th of that same year, when the expected count was 6.1. Together, they provide evidence of an anomalous increase of this type of crime report in the area during the first months of 2011.

\textbf{Real-time anomaly detection} In this study we have performed a retrospective analysis, however, to determine if a query is anomalous, only data that corresponds to the past of a given query is used, and therefore it mimics the real-time usage. Such implementation of this tool could aid in the discovery of emerging spatiotemporal anomalies, and in forecasting their future occurrences.




\section{Conclusion}
\label{conc}

We have proposed a way of analysing sexual violence through the assumption of an underlying latent structure. Such an assumption is sensible in cases where previous research has established latent causes of sexual violence. El Salvador is one of such cases but definitely not the only one. With that assumption in place, we have proposed the use of T-Cube data structure, in combination with statistical significance tests, to enable fast querying of the data and reliable discovery of anomalous spatiotemporal patterns. Using the framework we propose, we have analysed sexual violence data from El Salvador and we have found evidence of patterns that should be addressed by policy makers. The most salient of such patterns corresponds to the states in the East, where girls between 12 and 14 years old report their boyfriend as the perpetrator at rates that do not correspond to those in the rest of the country, with a peak taking place in the first half of 2008. Finally, we explain how such techniques could be implemented in real-time. The results presented in this paper demonstrate the ability of the proposed techniques to identify significant anomalies in this domain, indicating that it could potentially be used by policy makers for early detection of emerging patterns, which could enable development of effective policies and responses. 

\section*{Acknowledgements}
Carlos Eduardo Huertas and Suchit Chavez from Connectas, for their efforts to make the data publicly available and for sharing their knowledge about crime in El Salvador. This work has been partially funded by NSF grant 1320347 and by NIJ award 2013-IJ-CX-K007.

\bibliographystyle{apalike} 
\bibliography{bib.bib}

\begin{thebibliography}{}

\bibitem[Arana, 2005]{Ara2005}
Arana, A. (2005).
\newblock How the street gangs took central america.
\newblock {\em Foreign Affairs}, 84(3):98--110.

\bibitem[Berk et~al., 2005]{BHS2005}
Berk, R.~A., He, Y., and Sorenson, S.~B. (2005).
\newblock Developing a practical forecasting screener for domestic violence
  incidents.
\newblock {\em Evaluation Review}, 29(4):358--383.

\bibitem[Dubrawski, 2010]{DuZ2010}
Dubrawski, A. (2010).
\newblock The role of data aggregation in public health and food safety
  surveillance.
\newblock {\em Biosurveillance: Methods and Case Studies}, page 161.

\bibitem[Everitt, 1992]{Eve1992}
Everitt, B.~S. (1992).
\newblock {\em The analysis of contingency tables}.
\newblock CRC Press.

\bibitem[Gorr and Harries, 2003]{GoH2003}
Gorr, W. and Harries, R. (2003).
\newblock Introduction to crime forecasting.
\newblock {\em International Journal of Forecasting}, 19(4):551--555.

\bibitem[Hume, 2004]{Hum2004}
Hume, M. (2004).
\newblock “it’s as if you don’t know, because you don’t do anything
  about it”: gender and violence in el salvador.
\newblock {\em Environment and Urbanization}, 16(2):63--72.

\bibitem[Hume, 2007]{Hum2007}
Hume, M. (2007).
\newblock Mano dura: El salvador responds to gangs.
\newblock {\em Development in Practice}, 17(6):739--751.

\bibitem[J{\"u}tersonke et~al., 2009]{JMR2009}
J{\"u}tersonke, O., Muggah, R., and Rodgers, D. (2009).
\newblock Gangs, urban violence, and security interventions in central america.
\newblock {\em Security Dialogue}, 40(4-5):373--397.

\bibitem[Muggah, 2016]{Mug2016}
Muggah, R. (2016).
\newblock It's official: San salvador is the murder capital of the world.
\newblock \textit{LA Times}.

\bibitem[Neill and Gorr, 2007]{NeG2007}
Neill, D.~B. and Gorr, W.~L. (2007).
\newblock Detecting and preventing emerging epidemics of crime.
\newblock {\em Advances in Disease Surveillance}.

\bibitem[Planas, 2016]{Pla2016}
Planas, R. (2016).
\newblock How el salvador became the world’s most violent peacetime country.
\newblock \textit{Huffington Post}.

\bibitem[Ray et~al., 2007]{RMS2007}
Ray, S., Michalska, A., Sabhnani, M., Dubrawski, A., Baysek, M., Chen, L., and
  Ostlund, J. (2007).
\newblock T-cube web interface: a tool for immediate visualization, interactive
  manipulation and analysis of large sets of multivariate time series.
\newblock In {\em AMIA Annual Symposium proceedings}, pages 1106--1106.

\bibitem[Sabhnani et~al., 2007]{SMD2007}
Sabhnani, M., Moore, A., and Dubrawski, A. (2007).
\newblock Rapid processing of ad-hoc queries against large sets of time series.
\newblock {\em Advances in Disease Surveillance}, 2:66.

\bibitem[Secretariat, 2015]{Gen2015}
Secretariat, G.~D. (2015).
\newblock {\em Global Burden of Armed Violence 2015: Every Body Counts}.
\newblock Cambridge University Press.

\bibitem[Speizer et~al., 2008]{SGW2008}
Speizer, I.~S., Goodwin, M., Whittle, L., Clyde, M., and Rogers, J. (2008).
\newblock Dimensions of child sexual abuse before age 15 in three central
  american countries: Honduras, el salvador, and guatemala.
\newblock {\em Child abuse \& neglect}, 32(4):455--462.

\bibitem[Upton, 1992]{Upt1992}
Upton, G.~J. (1992).
\newblock Fisher's exact test.
\newblock {\em Journal of the Royal Statistical Society. Series A (Statistics
  in Society)}, pages 395--402.

\bibitem[Virginia López~Calvo, 2013]{CaP2013}
Virginia López~Calvo, G. S.~P. (2013).
\newblock El salvador: Truce for the gangs, no truce for women.
\newblock {Latin America Bureau}.

\bibitem[Waidyanatha et~al., 2010]{WSD2010}
Waidyanatha, N., Sampath, C., Dubrawski, A., Sabhnani, M., Chen, L., Ganesan,
  M., and Vincy, P. (2010).
\newblock T-cube web interface as a tool for detecting disease outbreaks in
  real-time: A pilot in india and sri lanka.
\newblock In {\em IEEE RIVF 2010}, pages 1--4. IEEE.

\bibitem[Wood, 2006]{Woo2006}
Wood, E.~J. (2006).
\newblock Variation in sexual violence during war.
\newblock {\em Politics \& Society}, 34(3):307--342.

\bibitem[Wood, 2009]{Woo2009}
Wood, E.~J. (2009).
\newblock Armed groups and sexual violence: when is wartime rape rare?
\newblock {\em Politics \& Society}, 37(1):131--161.

\end{thebibliography}
\end{document}